\documentstyle[prd,aps,epsf,graphicx]{revtex}

\DeclareFontFamily{OT1}{rsfs10}{}
\DeclareFontShape{OT1}{rsfs10}{m}{n}{ <-> rsfs10 }{}
\DeclareMathAlphabet{\mathscript}{OT1}{rsfs10}{m}{n}

\newcommand{\be}{\begin{equation}}
\newcommand{\ee}{\end{equation}}

\newcommand{\bea}{\begin{eqnarray}}
\newcommand{\eea}{\end{eqnarray}}
\newcommand{\ba}{\begin{array}}
\newcommand{\ea}{\end{array}}

\newcommand{\ns}{\normalsize}
\newcommand{\pt}{\partial}

\newcommand{\eqref}[1]{(\ref{#1})}

\def\a{\alpha}
\def\b{\beta}
\def\bt{\beta_{3}}
\def\bto{\beta_{30}}

\def\d{\delta}

\def\f{\phi}

\def\k{\kappa}

\def\m{\mu}

\def\p{\pi}

\def\rh{\hat{\rho}}

\def\t{\tau}

\def\bx{{\bf x}}
\def\bal{{\mbox{\boldmath $\alpha$}}}
\def\bq{{\bf q}}

\def\bd{{\bf d}}

\def\bp{{\bf p}}

\begin{document}

%%%%%%%%%%%%%%%%%%%%%%%%%%%%%%%%%%%%%%%%%%%%%%%%%%%%%%%%%%%%%%%%%%%%%%
\begin{titlepage}

\title{
\hfill{\ns DCPT-03/59\\}
\hfill{\ns hep-th/0312111 \\[1cm]}
{\LARGE Gauge Five Brane Dynamics And Small Instanton \\ Transitions In Heterotic Models}\\[1cm]}
\setcounter{footnote}{0}
\author{{\ns\large
 James Gray$^1$\footnote{email: j.a.gray2@durham.ac.uk},
\setcounter{footnote}{1}
 Andr\'e Lukas$^2$\footnote{email: a.lukas@sussex.ac.uk} and 
\setcounter{footnote}{2}
 Gavin I. Probert$^1$\footnote{email: g.i.probert@durham.ac.uk} \\[1cm]}
{\ns $^1$Centre for Particle Theory, University of Durham,}\\
      {\ns South Road, Durham DH1 3LE, United Kingdom.}\\[0.2cm]
      {\ns $^2$Department of Physics and Astronomy, University of Sussex}\\
      {\ns Falmer, Brighton BN1 9QH, United Kingdom.}\\[0.2cm]}

%\date{}

\maketitle

\vspace{1cm}

\begin{abstract}
We present the first examples of cosmological solutions to four-dimensional heterotic models which include an
evolving bundle modulus. The particular bundle modulus we consider corresponds to the width of a gauge five brane.
As such our solutions can be used to describe the evolution in one of these models after a small instanton transition. 
We find that certain properties are generic to these solutions, regardless of initial conditions. This enables us
to make some definite statements about the dynamics subsequent to a small instanton transition despite the fact that
we cannot microscopically describe the process itself. We also show that an effective description of the small instanton
transition by a continuous matching of fields and their first derivatives is precluded by the form of the respective 
low-energy theories before and after the transition.
\end{abstract}

\thispagestyle{empty}

\end{titlepage}

%%%%%%%%%%%%%%%%%%%%%%%%%%%%%%%%%%%%%%%%%%%%%%%%%%%%%%%%%%%%%%%%%%%%%%%

\section{Introduction}

A small instanton transition takes place when an M5 brane, moving in the orbifold direction of 
heterotic 
M-theory \cite{Horava:1996ma,Horava:1995qa,Witten:1996mz,Lukas:1997fg,Lukas:1998yy,Lukas:1998hk}, 
impinges upon one of the fixed planes \cite{Ganor:1996mu,Witten:1995gx}. 
It is thought that when such a collision occurs the M5 brane could
disappear and be replaced with a gauge five (G5) brane living upon the relevant fixed plane.

It is still not known whether small instanton transitions are dynamically allowed. By this we mean does the 
G5 brane, which initially appears with a width at the fundamental length scale, actually spread out with time to 
become an object describable by supergravity. Certainly we know that the process 
is consistent with conservation of various charges \cite{Buchbinder:2002ji}. 
We also know that the extra light states which appear as the M5 brane approaches the fixed point are the 
same as those which arise due to a G5 brane's core shrinking to zero size \cite{Ganor:1996mu,Witten:1995gx}. 
However, to determine 
whether dynamically this transition actually completes itself would require an understanding of the process 
at times when either the instanton is small when compared to the fundamental length scale or, if we approach the 
problem from the other side, when the M5 brane is within a fundamental length scale distance of the orbifold fixed plane.
This is an extremely difficult task due to the presence of a tensionless non-critical string which 
appears during the transition.

To make clear what we will achieve in this paper let us split a small instanton transition as described 
above into three distinct regimes.

\begin{itemize}
\item Firstly we have the regime where the M5 brane is moving in the bulk toward the orbifold fixed plane 
but has not yet come sufficiently close for the extra states associated with membranes attached to the
M5 and its orbifold mirror or the fixed point to become light. The dynamics associated with this regime have been 
described in detail in \cite{Copeland:2001zp,Copeland:2002fv}. 
However, even though this regime has been studied before we will find that, in order 
to describe the dynamics for the same compactification throughout, we need to derive some new 
solutions. This 
is because, as we will discuss in section II, the existing four dimensional moving five brane solutions \cite{Copeland:2001zp}
 are not
 compatible with being connected via a small instanton transition to solutions of 
 the known G5 brane effective actions \cite{Gray:2003vw}. We therefore derive some moving M5 brane solutions 
which are compatible with the solutions we shall present for regime three to illustrate that the qualitative behaviour as 
described in previously examined cases still holds here.

\item Secondly we have the regime of the small instanton transition itself where the tensionless non-critical 
string has to be taken into account. We shall have nothing to say directly about this regime in this paper.

\item Lastly we have the regime where the G5 brane size has dilated enough for the normal low energy 
supergravity description to be valid (i.e.~we are assuming that the transition is indeed dynamically allowed).
 The description of the dynamics associated with this regime is the main result of this paper. The 
evolution of the G5 brane after such a transition has not been investigated before. This is 
because the four dimensional theory describing the necessary degrees of freedom has only recently been derived
 \cite{Gray:2003vw}.
\end{itemize}

Even though we will not be able to say anything new here about the microscopic dynamics in the 
second of these regimes we will find that 
we can still say something concrete about what happens after a small instanton transition, under the 
assumption that it is indeed not dynamically excluded. The solutions which describe regime three
 have certain properties which are generic whatever choices of initial conditions 
are made for the moduli fields after collision. This allows 
us to make some qualitative statements about the dynamics after a small instanton transition despite the fact that we do 
not know how the solutions of regime one and three are matched at the point of collision and so we do not know what 
initial conditions to give to our solutions in regime three.

We will also show that the simplest possible effective description of the small instanton transition, corresponding to 
a continuous matching of the various fields and their first derivatives across regime two, is precluded. This is 
due to an incompatibility of this 
procedure with the Hamiltonian constraints on our solutions before and after the transition.

\vspace{0.5cm}
We would also like to stress that G5 branes can exist on the orbifold fixed planes regardless of 
whether or not there has been a small instanton transition in the past. As such the solutions we provide for 
regime three are interesting in their own right. In particular they constitute the first examples of 
cosmological solutions to heterotic models which include a gauge bundle modulus.

It may not be clear how the freely moving solutions we present here are compatible 
with recent work on moduli stabilisation 
in heterotic models \cite{stab}. We assume that all of the potentials we are 
neglecting are  compatible with the standard four dimensional heterotic 
effective theory obtained by reduction on the vacua of \cite{Lukas:1998yy,Lukas:1998tt}. Furthermore we 
require that the energy scales associated with the stabilisation potentials, for the values of moduli as 
given by our solutions, are several orders of magnitude smaller 
than the energy scale at which the four dimensional 
effective theory breaks down. We can then take the kinetic energies in our solutions to be large when 
compared to the 
potentials we are ignoring - thus justifying our approximation. One way in which this approximation 
can always be consistently achieved is to take the large radius limit.

\vspace{0.5cm}
The outline of this paper is as follows. In section II we give the effective actions which are necessary to 
describe the 
four dimensional physics before and after the small instanton transition. In section III we present the 
cosmological solutions associated with these actions. Section IV contains a discussion of what conclusions 
we can draw about the dynamics after a small instanton transition, an explanation of how we can rule out the simplest 
possible effective description of the small instanton transition and a 
brief outline of some possible generalisations of our solutions.

%%%%%%%%%%%%%%%%%%%%%%%%%%%%%%%%%%%%%%%%%%%%%%%%%%%%%%%%%%%%%%%%%%%%%%%%%%%%%%%%%%%%%%%%%%%%%%%%%%%%%%%%%%%%

\section{The four-dimensional effective actions}

M5 branes in the vacuum of heterotic M-theory have to be oriented in a certain way in order to preserve $N=1$ 
supersymmetry in the four dimensional theory \cite{Witten:1996mz}. Four of the five brane's six dimensions have 
to span the external 
Minkowski space. The remaining two world volume directions must then wrap a holomorphic curve within the 
Calabi-Yau. This leaves five dimensions in total which are transverse to the five brane. One of these is the orbifold 
direction and the remaining four lie within the Calabi-Yau.

The configuration of the G5 brane which appears after the collision is constrained by the requirement that a
 cohomology condition, required for consistency of the vacuum solution, 
should still be satisfied in the M5 brane's absence. 
This condition essentially encodes the need for the net magnetic charge on various compact manifolds to be zero.
 It is easy to see, using the expressions given in \cite{Witten:1996mz} for example, that these restrictions mean 
that the G5 brane after the collision must be wrapped in the same way as the M5 brane beforehand.

In the regime before the collision with the orbifold fixed plane the four dimensional action is known 
for any compactification manifold of $SU(3)$ holonomy and for the case of the five brane wrapping an 
arbitrary holomorphic curve within that manifold \cite{Derendinger:2000gy,Moore:2000fs,Brandle:2001ts,bandl}.

In the regime after the collision our knowledge is more restricted. The only known case (which also represents
the only known examples of kinetic terms for bundle moduli) was presented in \cite{Gray:2003vw}. For this 
result to be valid the compactification manifold is required to have one crucial property. Near the two 
cycle which the G5 brane wraps the manifold must be decomposable as a direct product of the two 
cycle and a four dimensional complex transverse space. A class 
of examples of such manifolds (which in addition 
should have $SU(3)$ holonomy of course) were provided  in \cite{Gray:2003vw}. These examples were based on Calabi-Yau 
threefolds constructed as the resolution of certain six dimensional orbifolds.

Given this restricted knowledge we should choose our compactification manifold and holomorphic 
curve in regime one such that we can write down the compatible theory, based upon the same compact manifold 
and holomorphic curve in regime three. 
Unfortunately the only four dimensional moving M5 brane solutions in the literature 
\cite{Copeland:2001zp} 
are for the case $h_{1,1}=1$ which is not compatible with this requirement. 
We will therefore derive a set of solutions for regime one, as well as regime 
three which, however, remains
 the case of primary interest in this paper. We shall see that all of the qualitative features of the moving M5 brane 
solutions in the $h_{1,1}=1$ case, \cite{Copeland:2001zp}, are reproduced for the solutions we shall 
present.

To be concrete we choose as our Calabi-Yau threefold a resolved $Z_8$ - I Coxeter orbifold with an $SO(5) 
\times SO(9)$ lattice \cite{Bailin:nk}. The homology class of the holomorphic 
curve wrapped by the five branes is then taken to be 
that associated with a wrapping of the subspace which undergoes 
identifications by the $SO(5)$ part of the lattice.

The appropriate four dimensional actions are then as follows. 
For regime three when the G5 brane is present we have \cite{Gray:2003vw},
\begin{equation}
\label{cmp_action_G5}
 S = 
\frac{1}{2\k_P^2}\int\sqrt{-g}\left[-R+\frac{1}{2}\pt_\m\f
     \pt^\m\f +\frac{1}{2}\pt_\m\bt\pt^\m\bt + \pt_\m\b\pt^\m\b+ 8q_{G5}e^{-\b}
     \pt_\m\rh\pt^\m\rh\right]\; . \label{S4G5}
\end{equation}
For regime one when the M5 brane is present we have \cite{Derendinger:2000gy,Moore:2000fs,Brandle:2001ts,bandl},
\begin{equation}
\label{cmp_action_M5}
 S = 
\frac{1}{2\k_P^2}\int\sqrt{-g}\left[-R+\frac{1}{2}\pt_\m\f
     \pt^\m\f +\frac{1}{2}\pt_\m\bt\pt^\m\bt + \pt_\m\b\pt^\m\b+ q_{M5}e^{\bt -\f}
     \pt_\m z\pt^\m z\right]\; . \label{S4M5}
\end{equation}

In these actions $\phi$, $\b$ and $\bt$ are moduli describing the compactification 
space (we have of course performed many consistent truncations 
from the full set of four dimensional fields). 
The modulus $\b$ determines the 
size of the four dimensional space transverse to our five branes in the Calabi-Yau from the perspective of the 
ten dimensional effective theory. 
The other Calabi-Yau modulus, $\bt$, determines the size of the two cycle the five branes wrap. 
The four dimensional dilaton is denoted by $\phi$. The kinetic terms for these moduli are the same in both 
cases because while the two different actions represent different physics they are based upon compactifications 
on the same internal manifold.

We also have the moduli fields that describe the relevant degrees of freedom of our five branes. The coordinate
 position of the M5 brane within the orbifold direction is denoted by $z$. The orbifold fixed planes are at $z=0$ and
$z=1$ in our coordinates. The width of the G5 brane in the first 
of the two actions is denoted by $\hat{\rho}$. In fact this is the coordinate size of the G5 brane 
where our conventions mean that the coordinate size of the transverse space in the Calabi-Yau is a constant,
$v_{\textnormal{trans}}^{\frac{1}{4}}$ . As such $\hat{\rho}$ can vary from $0$ up to 
$v_{\textnormal{trans}}^{\frac{1}{4}}$ as it expands in the transverse space. When plotting specific examples of our 
solutions we shall take $v_{\textnormal{trans}}^{\frac{1}{4}} =1$.

The factor of $e^{\beta_3}$ in the prefactor to the $z$ kinetic term in \eqref{cmp_action_M5} is a somewhat non-trivial 
consequence of the general expressions presented in \cite{Moore:2000fs,bandl} and the nature of the holomorphic curve which our M5 
brane is wrapping. This latter, as we have stated above, is determined by the need to match the configurations in regime one and 
three in a manner which is consistent with the cohomology condition and the action \eqref{cmp_action_G5}.

Now that we have written down the appropriate effective theories in the two regimes of interest we can 
proceed to find the cosmological solutions that form the core of the results of this paper.
%%%%%%%%%%%%%%%%%%%%%%%%%%%%%%%%%%%%%%%%%%%%%%%%%%%%%%%%%%%%%%%%%%%%%%%%%%%%%%%%%%%%%%%%%%%%%%%%%%%%%%%%%%%%%%

\section{Cosmological solutions including gauge-five brane or M-five brane moduli}

We wish to investigate the cosmological dynamics which follow from the actions 
\eqref{cmp_action_G5}
and \eqref{cmp_action_M5} presented in the
 previous section. In particular we are interested in the effects that a moving M5 brane or an 
expanding or contracting G5 brane can have on the evolution of the other four dimensional fields. 
The procedure for obtaining cosmological solutions from such actions is well known 
\cite{Lukas:1997iq}
 and so we shall not go into detail regarding this.

We make the ansatze,
\bea
ds^2 = - e^{2 \nu} d \tau^2 + e^{2 \alpha} d \bx^2 
\eea
where $\a$, $\b$, $\bt$, $\phi$, $\rh$ and $z$ are all functions of $\t$ only. 
The three-dimensional spatial submanifold is taken to be Ricci flat and we have either a 
$\hat{\rho}$ or a $z$ modulus depending on which case we are considering.

The equation of motion for the size (position) of the G5 (M5) brane can be trivially
integrated once.

\begin{eqnarray}
\dot{\hat{\rho}} &=& \tilde{s} \, e^{\beta +\nu  -3\a} \\
\dot{z} &=& \tilde{d} \, e^{\phi - \bt + \nu - 3\a}
\end{eqnarray}
Here $\tilde{s}$ and $\tilde{d}$ are arbitrary integration constants. 
These expressions can then be used in the remaining equations of motion to obtain a closed system in terms of
 $\alpha$, $\beta$, $\bt$ and $\phi$. The resulting equations can be written in the following compact form.
\bea
 \frac{d}{d\t}\left( EG\bal '\right)+E^{-1}\frac{\partial 
U}{\partial\bal}
  &=&0 \label{al_eom} \\ 
 \frac{1}{2}E{\bal '}^TG\bal '+E^{-1}U &=& 0\; , \label{N_eom}
\eea
The relevant four dimensional fields 
are arranged in a vector $\bal=(\alpha,\b, \bt, \f)$. The ``Einbein'' $E$ is defined as 
$E=e^{-\nu + \bd . \bal}$, where we have defined the vector 
$\bd=(3,0,0,0)$. The potential $U$ arises as a direct result of the presence of the dynamical 
G5 (M5) brane. 
It is determined by a constant $u$ and a vector $\bq$ which characterises the particular 
case under study. 

\begin{equation}
U = \frac{1}{2} u^2 e^{\bq . \bal}. \label{U}
\end{equation}

In the G5 brane case $u$ and $\bq$ take the following values.

\begin{eqnarray}
u^2 &=& 4 q_{G5} \tilde{s}^2 \\
\bq &=& (0,1,0,0)
\end{eqnarray}

In the M5 brane case they take a different form. This is due to the kinetic term for the brane's position modulus coupling 
differently to the other four dimensional fields.

\begin{eqnarray}
u^2 &=& \frac{1}{4} q_{M5} \tilde{d}^2 \\ \label{fin}
\bq &=& (0,0,-1,1)
\end{eqnarray}

In both cases the metric $G$, which is simply determined by the coefficients of the kinetic terms of
 the components of $\bal$ as given in the actions \eqref{cmp_action_G5}, \eqref{cmp_action_M5},
 takes the same form.

\begin{eqnarray}
G = \textrm{diag}(-3,\frac{1}{2},\frac{1}{4},\frac{1}{4})
\end{eqnarray}

By a judicious choice of gauge we can find the general solution to this system of 
equations in both cases. We shall now present these, starting with the G5 brane case.

\subsection{Solutions including gauge five-brane moduli}

We find the following solutions describing the evolution of the gauge five brane's size modulus and the other 
four dimensional fields. Our solutions are presented in comoving gauge for ease of physical interpretation.

\bea
\label{G51}
 \a &=& \frac{1}{3}\ln\left|\frac{t-t_0}{T}\right|+\a_0 \\ \label{evil3}
 \b &=& p_{\b ,i}\ln\left|\frac{t-t_0}{T}\right|+
        (p_{\b ,f}-p_{\b ,i})\ln\left(\left|\frac{t-t_0}{T}\right|^{-\d}+1
        \right)^{-\frac{1}{\d}}+ \b_0\\
 \bt &=& p_{\bt,i}\ln\left|\frac{t-t_0}{T}\right|+\bto \\
 \f &=& p_{\f,i}\ln\left|\frac{t-t_0}{T}\right|+\f_0 \\ \label{rhoeqn}
\rh &=& \hat{s} \, \left( 1 + \left| \frac{T}{t- t_0} \right|^{\delta} \right)^{-1} 
+ \rh_0 \; 
\eea

It is reassuring to note that the expansion power for the four dimensional scale factor is 1/3, 
as expected for kinetic energy driven 
expansion in the Einstein 
frame. The constants $p_{\b,n}$, $p_{\b 3,n}$ and $p_{\phi,n}$ are subject to the constraint,
\begin{equation}
2p^{2}_{\b ,n} + p^{2}_{\bt ,n} + p^{2}_{\f ,n} = \frac{4}{3}\; , \label{G5con}
\end{equation}
for $n=i,f$. In this expression $p_{\bt,f}= p_{\bt,i}$ and $p_{\phi,f}= p_{\phi,i}$. We shall see the physical meaning of \eqref{G5con} 
shortly.
The constants, $\b_0$ and $\hat{s}$ are subject to a constraint.

\begin{eqnarray}
\label{evil}
\b_0 = \ln(2 q_{G5} \hat{s}^2)
\end{eqnarray}
The constant $\delta$ takes the following form.

\begin{equation}
\label{evil2}
\delta = -p_{\b ,i} 
\end{equation}
Finally we have the arbitrary integration constants $\hat{\rho}_0$ ,$\a_0$, $\b_{30}$, $\phi_0$, $t_0$ and $T$.

The above class represents the first examples of cosmological solutions to heterotic models which include a gauge 
bundle modulus. As such they are of interest in their own right as well as in the more specific context we are
 considering in this paper. They detail some of the kind of effects we can expect a dynamical gauge bundle to have 
on the cosmological evolution of the theory.
\vspace{0.5cm}

We shall now spend some time describing in detail the physics which these solutions represent. The 
broad features presented may be familiar to some readers as formal analogies can be drawn between 
the system of current interest and systems involving potentials due to form fields \cite{Lukas:1997iq} or 
moving M5 branes \cite{Copeland:2001zp}.

We first consider the limitations placed on the range of t in the above solutions by the 
requirement that the logarithms remain well defined. We find two possible allowed ranges.

\be
 t\in\left\{\ba{clll}
       \left[ -\infty ,t_0\right]\;
,&(-)\;{\rm 
branch} \\
       \left[ t_0,+\infty\right]\; 
,&(+)\;{\rm branch}
       \ea\right.\; .
\ee

Thus we find a separation of our cosmological solutions into so called positive and negative time branches. 
The negative time branch starts with an asymptotically flat universe and evolves into a curvature 
singularity at $t=t_0$ whereas the positive time branch starts at a curvature singularity at time $t_0$ 
and evolves to less and less highly curved configurations with time.
In this paper we shall focus, in our discussions, on positive time branch solutions. The reader who wishes 
to consider the negative time branch case need only take the time reverse of our analysis.

There is a redundancy in the constants labelling our solutions. This is due to a symmetry under the following 
transformations.

\bea
\delta &\rightarrow& -\delta \\
\bp_i &\rightarrow& \bp_f \\
\bp_f &\rightarrow& \bp_i \\
\hat{s} &\rightarrow& - \hat{s} \\
\hat{\rho}_0 &\rightarrow& \hat{\rho}_0 + \hat{s}
\eea
The physical situation described is unchanged under this change of constants. To avoid over counting solutions 
we shall therefore employ the following convention.

\bea
\label{evillast}
\delta < 0  \;\;\; (+) \textnormal{branch} \\
\delta > 0  \;\;\; (-) \textnormal{branch}
\eea

Now let us consider how the various fields evolve in time. 
It can be seen from equation \eqref{rhoeqn} that at either end of the t range, as $t \rightarrow t_0$ or as 
$t \rightarrow \infty$, $\hat{\rho}$ approaches a constant. Now $\hat{\rho} = e^{-\frac{\beta}{2}} \rho$ 
where $\rho$ is the physical width of the gauge five brane. 
The other factor in $\hat{\rho}$, $e^{\frac{\beta}{2}}$,  
measures physical 
lengths in the 
transverse space to the G5 brane in units of a fixed coordinate size for that submanifold. 
Thus $\hat{\rho}$ being a 
constant has the physical meaning that the gauge five brane size is scaling with the size of the transverse portion 
of the
compactification manifold. Any change in the soliton's width is due to an overall change in the size of 
the transverse space and not a change in size of the object relative to its surroundings.

While the G5 brane is of 'constant size' in this manner its kinetic term drops out of the effective 
four dimensional theory \eqref{cmp_action_G5}. Since this kinetic term was the only indication from our four dimensional 
perspective that the gauge five brane was present (we are not interested here in four dimensional gauge 
groups etc.) this means that asymptotically, when the size is constant, the solutions should approach 
those that have been obtained in the study of four dimensional heterotic M-theory in the absence of gauge 
five brane moduli \cite{Brandle:2000qp}.

An examination of the solutions \eqref{G51} - \eqref{rhoeqn} shows that this is indeed the case. We find that 
when $t \rightarrow t_0$ we have a so called rolling radius solution with expansion powers for the geometrical 
moduli
given by $\bp_i$, whereas when $t \rightarrow \infty$ we have another, in general different, such solution 
characterised by expansion powers $\bp_f$. As is usual these expansion powers are subject to a constraint 
\eqref{G5con}.

At times intermediate between the two asymptotic regions described above it is clear that $\hat{\rho}$, 
and so the size of the five brane relative to the compactification space, changes with time. In fact 
the size of the soliton monotonically changes from its initial value, $\hat{s} + \hat{\rho}_0$, to its final value, 
$\hat{\rho}_0$,
at some time determined 
by $T$. As it does so it affects, via the non-trivial coupling of its kinetic 
term to $\beta$, 
the other moduli, which can be seen to no longer take the form of a simple rolling radius solution at these times. 
In fact, equation
 \eqref{evil3} tells us that the five brane motion maps the initial rolling radius solution into a final one determined 
by the mapping given below.

\begin{equation}
 \left(\ba{c}p_{\b ,f}\\p_{\bt ,f}\\p_{\f ,f}\ea\right) = P
 \left(\ba{c}p_{\b ,i}\\p_{\bt ,i}\\p_{\f ,i}\ea\right)\; ,\qquad
 P = \left(\ba{rrr}-1&0&0\\0&1&0\\0&0&1\ea\right)\; . \label{map1}
\end{equation}

In particular the final 
set of expansion powers is completely determined by the initial one. The mapping given 
above is its own inverse. This is simply a consequence of the time reversal symmetry which exists in our 
four dimensional effective action \eqref{cmp_action_G5}.

It should be noted that, if we are in a positive time branch for example, not all of the possible rolling 
radius solutions are available to us as initial configurations at $t \rightarrow t_0$.
The rolling radii configurations which are 
allowed as initial states for solutions with changing G5 brane width are those with $p_{\beta ,i}>0$ (in the
 positive time branch case). In other 
words the transverse space to the G5 brane within the Calabi-Yau is initially expanding and collapses at late times.

There are a few special solutions in the class presented here for which $\hat{\rho}$ is constant
throughout the evolution. These correspond to solutions where the transverse space is constant 
in size, $p_{\b,i} = 0$. Such configurations 
are indistinguishable in four dimensions from situations in which the G5 brane is not present.

\vspace{0.5cm}

We should at this point, make a few comments about when our four dimensional theory, and so the solutions 
we have derived in this section, is valid. 
We of course have all of the usual restrictions on the validity of the four dimensional effective description of 
heterotic M-theory. We also have an additional restriction in that the
theory is also not valid when the G5 brane's width becomes comparable to the size of the transverse 
space. This is due to an approximation that was made in obtaining the four dimensional theory \cite{Gray:2003vw}.

Let us consider in detail the requirement that the extra dimensions should be larger than the fundamental scale. 
Consider the 
constraint on the integration constants \eqref{evil}, the value of $\delta$ \eqref{evil2} and the expansion power 
$p_{\beta,f}$ in terms of the initial expansion powers \eqref{map1}. If we use all of this information in equation 
\eqref{evil3} for $\beta$ we obtain the following.

\bea
\label{evil4}
e^{\beta} = \frac{2 \a' (2 \pi)^2 \hat{s}^2}{v_{trans}} \left[ \left|\frac{t-t_0}{T}\right|^{p_{\beta i}} \left( 
\left|\frac{t-t_0}{T}\right|^{p_{\beta i}} + 1 \right)^{-2} \right]
\eea
The function in the square brackets in equation \eqref{evil4} is alway less than 1. 

For the length scale 
associated with the transverse space to be much greater than the fundamental scale we require,

\bea
\label{evil5}
e^{\beta} \sqrt{v_{trans}} >> \a' 
\eea
We see from equation \eqref{evil4} that the only way to achieve this is to have $\hat{s}>>v_{trans}^{
\frac{1}{4}}$.
 In other words we need the total change in size of the G5 brane during the evolution to be orders of 
magnitude greater than the size of the compact space in which it lives. This, of course, means that the full solutions 
with all of the behaviour described above are not valid in any one particular case. We can only follow the evolution 
for a certain period while the size of the G5 brane is bigger than the fundamental scale and smaller than the size of 
the compactification space. It should be stressed however that by a judicious choice 
of the constant $\hat{\rho}_0$ any region of the solutions illustrated above can fall within this describable range. We just 
do not have valid solutions where we can follow the dynamics through from one asymptotic regime to the other.

It should be noted that because the transverse space always collapses asymptotically in these solutions our 
description always breaks down at late times.

\begin{figure}[ht]\centering
\includegraphics[height=7.9cm,width=10cm, angle=0]{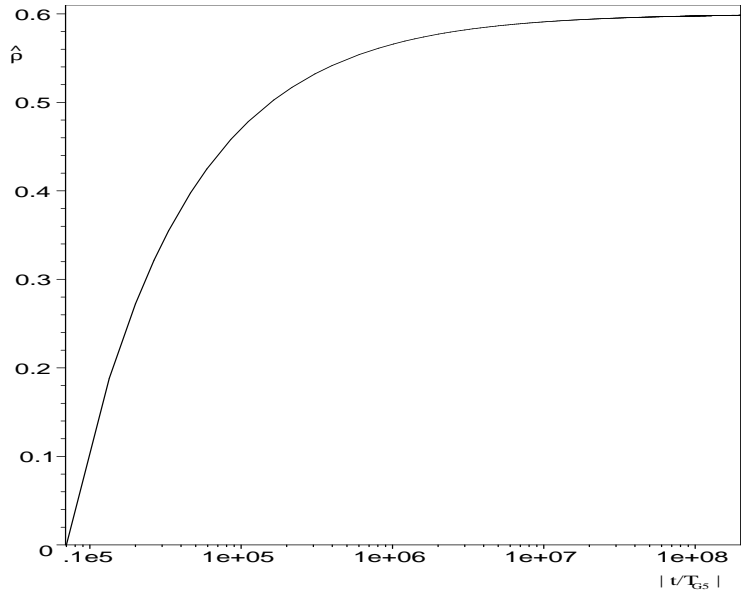}
        \caption{\emph{Plot of the coordinate size of a gauge five brane including the regime where our four dimensional
 effective theory is valid. We have chosen the set up $T_{G5}=1$, $\hat{s}=-100$ and $\hat{\rho}_0=0.6$, with 
$p_{\beta,i}=p_{\bt,i}=p_{\f,i}=\sqrt{\frac{1}{3}}$, $\beta_{0}$ given by \eqref{evil}, $\bto=5$ and $\f_{0}=10$, giving 
$\delta=-\sqrt{\frac{1}{3}}$. The G5 brane appears after a small instanton transition with essentially zero size. It then expands,
 asymptotically approaching a constant size which is 
smaller than the length scale associated with the transverse space. This type of evolution 
occurs in certain, fairly special, examples of our solutions as described in section IV. }}
\label{G5a}
\end{figure}

\begin{figure}[ht]\centering
\includegraphics[height=7.9cm,width=10cm, angle=0]{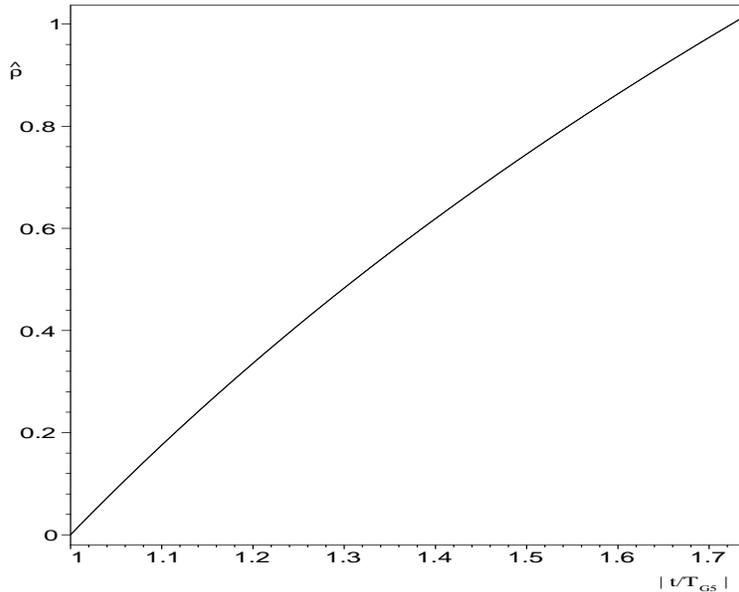}
        \caption{\emph{Plot of the coordinate size of a gauge five brane including the regime where our four dimensional
 effective theory is valid. We have chosen the set up $T_{G5}=0.21$, $\hat{s}=-17.28$ and $\hat{\rho}_0=8.61$, with $p_{\beta,i}=0.43$, 
$p_{\bt,i}=-0.45$,
$p_{\f,i}=0.87$, $\beta_{0}=6.39$, $\bto=17.98$, $\f_{0}=19.28$ and $\a_0 = -0.51$. The G5 brane appears after a small instanton
 transition with essentially zero size. It then expands rapidly, quickly reaching a size comparable to that of the transverse space. 
At this point our four dimensional theory breaks down and so we have truncated our plot appropriately. This type of evolution is fairly 
generic for the class of solutions we have presented here as described in section IV.}}
\label{G5c}
\end{figure}

Examples of the dynamics of the bundle moduli that we have obtained here are plotted, including the regimes where our four 
dimensional description 
is valid, in figures \ref{G5a} and \ref{G5c}.

\vskip 0.4cm

%%%%%%%%%%%%%%%%%%%%%%%%%%%%%%%%%%%%%%%%%%%%%%%%%%%%%%%%%%%%%%%%%%%%%%%%%%%%%%%%%%%%%%%%%%%%%%%%%%%%%%%%%%%%%%%%%%%%%

\subsection{Solutions including M-five brane moduli}

Our goal in presenting new solutions for moving M5 branes in heterotic M-theory in this section is simply 
to show that the qualitative results found elsewhere \cite{Copeland:2001zp} 
also apply to the current case. As such we shall be 
quite brief in our exposition. The explicit solutions, in comoving gauge,
for the case of an M5 brane traversing the bulk are the following.

\bea
\label{m51}
 \a &=& \frac{1}{3}\ln\left|\frac{t-t_0}{T}\right|+\a_0 \\
 \b &=& p_{\b,i}\ln\left|\frac{t-t_0}{T}\right|+\b_0 \\ \label{m5b}
 \bt &=& p_{\bt ,i}\ln\left|\frac{t-t_0}{T}\right|+
        (p_{\bt ,f}-p_{\bt ,i})\ln\left(\left|\frac{t-t_0}{T}\right|^{-\d}+1
        \right)^{-\frac{1}{\d}}+\bto \\ \label{m5f}
 \f &=& p_{\f ,i}\ln\left|\frac{t-t_0}{T}\right|+
        (p_{\f ,f}-p_{\f ,i})\ln\left(\left|\frac{t-t_0}{T}\right|^{-\d}+1
        \right)^{-\frac{1}{\d}}+\f_0 \\ \label{m55}
 z &=& d \, \left( 1 + \left| \frac{T}{t- t_0} \right|^{\delta} \right)^{-1} 
+ z_0 \; .
\eea

The expansion powers in this solution are subject to the following constraint.

\begin{equation}
\label{constraint}
2p^{2}_{\b ,n} + p^{2}_{\bt ,n} + p^{2}_{\f ,n} = \frac{4}{3}\;  
\end{equation}
for $n=i,f$. In this expression $p_{\b,f} = p_{\b,i}$.

In addition to these expansion powers the solution depends upon the arbitrary constants $T$ and
$t_0$. There are a further set of constants $\a_0$, $\b_0$, $\b_{30}$ and $\phi_0$ which are also 
subject to a constraint.

\begin{equation}
\label{scalconstr}
\f_0 - \b_{30} =\ln \left(\frac{1}{2} d^{2}q_{M5}\right) \; .
\end{equation}
The constant $\delta$ is determined by a particular combination of the expansion powers.
\begin{equation}
\delta = p_{\bt ,i} - p_{\f ,i} \; .
\end{equation}
Finally we have the two constants associated with the integration of the five brane's position 
modulus equation of motion, $d$ and $z_0$.

The solutions given in equations \eqref{m51} - \eqref{m55} describe the following physical situation. They again 
take the form of either positive or negative time branches and, as in the G5 brane case, we take 
$\delta>0$ in the (-) branch and $\delta<0$ in the (+) one. The system starts 
out in a rolling radius solution associated with the theory in the absence of an M5 brane \cite{Brandle:2000qp}.
 The M5 brane is initially stationary. Then, at some time determined by T and $t_0$, the M5 
brane moves in the orbifold direction until it comes to rest at some new position. During this
 motion the five brane, through its coupling to the four dimensional metric moduli, changes the 
 initial rolling radius solution into a different one. Thus the system finishes up in some different 
final rolling radius solution with the five brane again at rest. We can write
 down a mapping which describes how the moving brane takes one set of expansion powers into another as follows.
\begin{equation}
\label{mapM5}
 \left(\ba{c}p_{\b ,f}\\p_{\bt ,f}\\p_{\f ,f}\ea\right) = P
 \left(\ba{c}p_{\b ,i}\\p_{\bt ,i}\\p_{\f ,i}\ea\right)\; ,\qquad
 P = \left(\ba{rrr}1&0&0\\0&0&1\\0&1&0\ea\right)\; 
\end{equation}

\begin{figure}[ht]\centering
\includegraphics[height=9cm,width=11cm, angle=0]{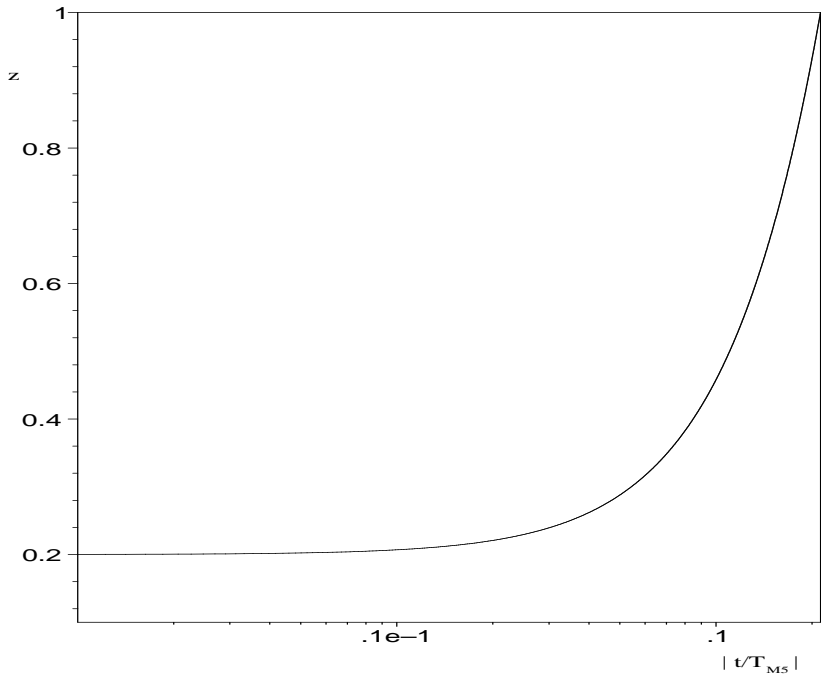}
        \caption{\emph{Plot of the coordinate position of an M5 brane including the regime where our four dimensional
 effective description is valid. We have chosen the set up $T_{M5}=1$, $d=-10$ and $z_0=10.2$,
 with $p_{\beta,i}=0$, $p_{\bt,i}=-\sqrt{\frac{1}{3}}$, $p_{\f,i}=1$, $\beta_{0}=5$, $\bto=17$, $\a_0 = 0$,
 and $\f_{0}$ determined by \eqref{scalconstr}. The M5 brane starts out close to the left boundary plane at z=0 and moves across the 
orbifold direction to collide with the right boundary at z=1.}}
\label{M5a}
\end{figure}

It should be noted that one of the conditions for small warping in the heterotic M-theory vacuum \cite{Lukas:1998tt} is always violated
 asymptotically for these solutions (assuming that the five brane moves). In fact it is important to emphasise that in 
order to keep this warping parameter much less than one in any region of the solution it is necessary that the coordinate 
distance that the M5 brane moves must be much larger than the size of the orbifold. This arises in a very similar manner 
to the constraint on $\hat{s}$ in the previous section. It means that here too we cannot have a single solution which 
has all of the asymptotic features we have described in a single regime for which the four dimensional theory is valid. 
An example of the dynamics of a moving five brane as described by our solutions is provided in figure \ref{M5a}
 including the portion of the time 
evolution where the four dimensional theory is valid.

In short we find that these solutions do indeed share all of the qualitative properties of the solutions 
presented previously for the $h_{1,1}=1$ case \cite{Copeland:2001zp} 
(although the restriction mentioned above was not made explicit in that paper). 
These solutions have been described in some detail in the 
literature and the interested reader is referred there for further discussion.

\vskip 0.4cm

%%%%%%%%%%%%%%%%%%%%%%%%%%%%%%%%%%%%%%%%%%%%%%%%%%%%%%%%%%%%%%%%%%%%%%%%%%%%%%%%%%%%%%%%

\section{Conclusions about possible dynamics after a small instanton transition and generalisations of our results}

Given the results of the previous section, although we cannot say whether or not the small instanton transition
 is dynamically allowed, we can make some definite statements about the dynamics of the five branes 
before and after such a transition.

The dynamics associated with the M5 brane before the collision is the same as would have been expected from 
analogy with previous work \cite{Copeland:2001zp}. Consider a case where
 the M5 brane starts out at some, approximately constant, position in the bulk with 
the system evolving in some rolling radius solution. Then at some time determined by the integration 
constants of the system ($T$ and $t_0$) the M5 brane starts to move as described in equation \eqref{m55}. 
This causes the other moduli to depart from their 
'rolling radius' behaviour as described in equations \eqref{m5b} and \eqref{m5f}. 
The M5 brane moves through the bulk and impacts upon one of the orbifold fixed planes. The fixed plane that is hit can be determined by 
 an appropriate choice of the constants $d$ and $z_0$. This impact marks the end of regime 
one in our parlance.

There is then a regime which is poorly understood. The tensionless non-critical string comes into play and 
the dynamical degrees of freedom some how switch over, we postulate, from those associated with an M5
 brane to those associated with the G5 brane. If this G5 brane reaches a width which is 
larger than the fundamental scale we then enter the third of our regimes.

\vspace{0.3cm}

In this third regime the soliton appears with fundamental length scale width. It then expands
 monotonically with time in one of two possible ways. 

\vspace{0.2cm}

For generically acceptable choices of the integration constants,
specifically for most choices which result in the size of the transverse space being much bigger than the fundamental 
scale, the G5 brane rapidly and monotonically increases in size until it reaches a size comparable to that of the 
compact manifold it lives in. This can be seen by looking at values of $\hat{s}$ which are compatible with the constraint as 
given in
\eqref{evil4} and \eqref{evil5} and examining the impact that such choices of integration constant have on the G5 brane 
evolution via equation \eqref{rhoeqn}.
  Once the G5 brane is comparable in size to the subspace of the Calabi-Yau transverse to it
 we can no longer trust our four dimensional theory. 
It is interesting that we are able 
to say something so specific about the evolution after a small instanton transition, that the gauge field spreads 
out very rapidly from being a localised lump to being a diffuse configuration. This is also a desirable result from the 
point of view of the phenomenological models based upon small instanton transitions  \cite{Khoury:2001wf,Bastero-Gil:2002hs}. 
This is because
 it means that, in a regime where 
non-perturbative potentials are unimportant, the G5 brane does not shrink back to zero size again once 
it has been created and the transition, once it occurs, is permanent. 
The rapid expansion of the G5 brane induces a 
complicated evolution in the modulus describing the size of the space transverse to it within the Calabi-Yau,
 as described in \eqref{evil3}.

\vspace{0.2cm}

For certain choices of integration constant however the G5 brane
 will expand, again from fundamental scale size, to reach a constant width 
which is smaller than the transverse space. By constant size
 we mean that the objects coordinate size, $\hat{\rho}$, will remain constant, i.e.~the ratio of its size to 
that of the transverse space is fixed. After this the G5 
brane will simply scale in size with the transverse space and the four dimensional metric moduli will 
settle down into some new rolling radius solution. In fact the choice of integration constants required to obtain this 
behaviour is fairly contrived. If we 
want the transverse space to the G5 brane in the Calabi-Yau to be at least several orders of magnitude larger than the fundamental scale 
at its largest then, from equation \eqref{evil4}, we must take $\hat{s} \geq 100 v_{\textnormal{trans}}^{\frac{1}{4}}$. 
In our conventions however the coordinate size of the transverse space is merely of order $v_{\textnormal{trans}}^{\frac{1}{4}}$. 
Looking at equation \eqref{rhoeqn} we see then that for the final size of the G5 brane, $\hat{\rho}_0$, to be smaller than the 
size of the transverse space we require $\hat{\rho}_0 / \hat{s} \leq 0.01$.

This possible late time behaviour, with the G5 brane degrees of freedom dropping out of the dynamics,
 gives some justification to the approach that was taken in several recent papers 
\cite{Khoury:2001wf,Bastero-Gil:2002hs} of ignoring the 
soliton's moduli after the collision. This approach was taken because, at the time,
the four dimensional theory including G5 brane moduli was not known. Nevertheless 
we see that, providing we are looking at the system at times long enough after 
the collision,
this approach turns out to give a very reasonable approximation to the true physical situation for certain 
special configurations. In general, however, we should include one of the more generic solutions described above in our analysis of 
these scenarios. 
The problem then is that we loose control of the four dimensional effective theory very quickly. This however, is the 
best that can be done with current technology, and at least it is possible to describe the dynamics on either side of 
the small instanton transition and show that the G5 rapidly spreads out from its initially small size.

Another general comment we can make is that the size of the transverse space to the G5 brane within the Calabi-Yau always 
ends up collapsing after a small instanton transition. This follows from equations \eqref{evil3}, \eqref{evil2} and \eqref{evillast} 
and the requirement that the G5 brane grows from its very small initial size so that we can describe it using supergravity.
This last means that when our description of regime three begins the G5 brane modulus is increasing and so we 
cannot pick a solution of constant width. Presumably this collapse is halted by whatever 
mechanism finally stabilises the extra dimensions (subject to the assumptions we have made about this mechanism here 
as detailed in the introduction).

To recap, the crucial point is that, due to some generic properties of our evolving G5 brane solutions, we can
 make some qualitative comments about the dynamics after a small instanton transition
despite the fact that, because of our ignorance of regime two, we have no rigorous procedure for matching
 the solutions in regime one to those in regime three across the transition.

\subsection{Consequences of possible sets of matching conditions.}

Although we have no microscopic description of the dynamics of regime two we can still make an educated guess as to how to match our
solutions for regime one and three across the transition. In many diverse situations where extra light states appear
one can match across the transition simply by assuming that the moduli and scale 
factor are smooth and continuous throughout. One recently studied example is the flop transition \cite{Brandle:2002fa}.

It is easy to see that this procedure is in fact not possible in this case, if we assume that $z$ and $\hat{\rho}$ each describe part 
of one and the same
 continuous flat direction in moduli space. If we match the values and derivatives of $\a, \b, \bt$ and 
$\phi$ at the collision then we find that $\dot{\hat{\rho}}$ after the collision is determined by this data and the constraint 
\eqref{N_eom}. 
It 
turns out that, due to the different prefactors of the two five brane moduli in actions \eqref{cmp_action_G5} and \eqref{cmp_action_M5}, 
that the value we obtain 
in this manner for $\dot{\hat{\rho}}$ after the collision is not the same as the value of $\dot{z}$ beforehand.

In each of the cases where this naive matching of fields and their derivatives has been shown to be a good approximation 
the 
extra light states which have appeared have been particles. In  our case the extra light states are associated with a tensionless string.
 This difference, along with the possibility that $\hat{\rho}$ and $z$ do not constitute part of the same flat direction,
  could be responsible for the fact that the simple matching procedure that has worked in the past does 
not work in this case. It is interesting that we are 
able to show so simply and explicitly that the usual procedure fails. In addition, 
whatever the precise dynamics of the small instanton transition, we conclude that it 
has to somehow account for the required discontinuity in the fields described above.

The next most simple set of matching conditions would be to match all of the metric 
moduli, the four dimensional scale factor and these fields' 
derivatives as before and then to determine the bundle modulus' derivative at collision as described in the previous 
paragraphs. This, combined with the fact that the G5 brane starts out at 'zero' size is enough to specify 
a complete set of initial conditions for regime three, given a solution for the moving M5 brane in regime one.
 It is of interest to see if we can add anything to our general analysis, as presented at the start of this section, if we 
assume that we can pair up solutions in this particular manner. In fact the data plotted in figures \ref{M5a} and 
\ref{G5c} has been 
chosen such that the two can be matched together in this way.

Once we have followed this procedure we can describe how the G5 brane evolves after the collision given how the M5 brane 
was moving before hand. We can use this to elaborate on our expectation, as mentioned above, that the G5 brane will generically evolve 
as shown in figure \ref{G5c} as opposed to as shown in figure \ref{G5a}.
Generically we can see from \eqref{m55} that we expect the M5 brane to be moving rapidly at collision. This is essentially because our 
solution for $z$ is an 'S' shaped curve implying that the M5 brane, except in two small regimes, is moving rapidly or not at all. 
Because the M5 brane obviously has to be moving at collision this means that we expect it to be on the steep part of the 'S' when we 
enter regime two. We may then consider what this means for the resulting initial rate of change of $\hat{\rho}$ after the collision. 
The bundle modulus moves in a similar way, also forming an 'S' curve when the complete solution (of which we can physically sample but 
part) is plotted. By examining the constraints \eqref{N_eom}-\eqref{fin} 
we can see that if we have a large kinetic energy associated 
with the M5 brane before collision we would also expect a large kinetic energy for the bundle modulus after collision. In other words 
we would expect to find our solution in regime three on the steep part of the curve. This adds to and is in accord with
comments we have made at the start 
of this section where we were not considering a particular form of matching.

The plots \ref{M5a} and \ref{G5c} for $z$ and $\hat{\rho}$ show that, in a particular such example, 
the M5 brane is indeed moving rapidly at collision and that this results in a rapidly expanding G5 
brane in the final regime.

\subsection{Generalisations of our results.}

One possible direction for future research is to include more moduli in our cosmological analysis. We have,
 in this paper, just included the key moduli for the phenomena we are trying to investigate. There are many other 
interesting moduli which we could include, for example the moduli which delineate the holomorphic curve 
which the five branes wrap. We could also generalise our cosmological ansatz to include spatial curvature or,
perhaps more interestingly, an ideal gas living on either of the fixed planes. These last two possibilities, for example,
could even be solved completely analytically using the Toda theory methods described in \cite{Lukas:1997iq}.

Often in freely moving moduli solutions such as those presented in this paper
 we can use certain symmetries of the action to generate 
solutions involving axion fields from solutions which only include the real parts of the various superfields.
 It is well known \cite{Copeland:2001zp} that it is no longer possible to include all of the axions in this manner 
when an M5 
brane is present due to the object explicitly breaking some of the relevant symmetry. The same is 
in fact true in the presence of a gauge five brane and so, due to the numerous cross terms in the relevant 
action \cite{Gray:2003vw}, including all of the axions in these solutions would be a difficult challenge.

Finally it would be interesting to try and include the effects on our solutions of the 
potentials which are present due to
gaugino condensation, flux on the internal manifold, and membrane instanton effects \cite{stab}. We would expect such solutions to 
agree with those presented here when the kinetic energy of the moduli fields is sufficiently large (as described in the introduction) 
but deviate significantly, for example, at late times if we have not reached a large radius limit.

%%%%%%%%%%%%%%%%%%%%%%%%%%%%%%%%%%%%%%%%%%%%%%%%%%%%%%%%%%%%%%%%%%%%%%%%%%%%%%%%%%%%%%%%%%%%%%%%%%%%%%%%%%%%

\section*{Acknowledgements}
JG is supported by a PPARC postdoctoral rolling grant, AL by a PPARC advanced fellowship and GIP by a PPARC 
studentship.

%%%%%%%%%%%%%%%%%%%%%%%%%%%%%%%%%%%%%%%%%%%%%%%%%%%%%%%%%%%%%%%%%%%%%%%%%%%%%%%%%%%%%%%%%%%%%%%%%%%%%%%%%%%%

\end{document}